# Near-field GHz rotation and sensing with an optically levitated nanodumbbell

Peng Ju[1], Yuanbin Jin[1], Kunhong Shen[1], Yao Duan[2], Zhujing Xu[1], Xingyu Gao[1], Xinjie Ni[2], Tongcang Li[1,3,4,5]*

[1]Department of Physics and Astronomy, Purdue University, West Lafayette, Indiana 47907, USA.
[2]Department of Electrical Engineering, The Pennsylvania State University, University Park, Pennsylvania 16802, USA.
[3]Elmore Family School of Electrical and Computer Engineering, Purdue University, West Lafayette, Indiana 47907, USA.
[4]Birck Nanotechnology Center, Purdue University, West Lafayette, Indiana 47907, USA.
[5]Purdue Quantum Science and Engineering Institute, Purdue University, West Lafayette, Indiana 47907, USA.
*Corresponding author. Email: tcli@purdue.edu

**Abstract**

A levitated non-spherical nanoparticle in a vacuum is ideal for studying quantum rotations and is an extremely sensitive torque and force detector. It has been proposed to probe fundamental particle-surface interactions such as the Casimir torque and the rotational quantum vacuum friction, which require it to be driven to rotate near a surface at sub-micrometer separations. Here, we optically levitate a silica nanodumbbell in a vacuum at about 430 nm away from a sapphire surface and drive it to rotate at GHz frequencies. The relative linear speed between the tip of the nanodumbbell and the surface reaches 1.4 km/s at a sub-micrometer separation. The rotating nanodumbbell near the surface demonstrates a torque sensitivity of $(5.0 \pm 1.1) \times 10^{-26}$ NmHz$^{-1/2}$ at room temperature. Moreover, we levitate a nanodumbbell near a gold nanograting and use it to probe the near-field intensity distribution beyond the optical diffraction limit. Our numerical simulation shows it is promising to detect the Casimir torque between a nanodumbbell and a nanograting.

**Introduction**

Recently, quantum ground state cooling of the center-of-mass (CoM) motion of an optically levitated nanosphere in a vacuum was achieved (*1-4*), showing the great potential of levitated nanoparticles for studying macroscopic quantum mechanics (*5, 6*). Meanwhile, levitated dielectric particles in a vacuum are ultrasensitive force detectors (*7, 8*), and their CoM motion has been proposed to detect short-range forces (*9-15*), dark matter (*16*), dark energy (*17*), high-frequency gravitational wave (*18, 19*), and quantum effects in gravity (*20, 21*). Besides the CoM motion (*22-24*), there is increasing interest in the torsional (*25, 26*) and rotational motions (*27-32*) of birefringent (*27, 29*) and non-spherical particles (*33-36*). Levitated non-spherical nanoparticles in free space have been driven to rotate at GHz frequencies (*36-39*), and cooled by active feedback (*40, 41*), coherent scattering (*42, 43*), and spin-optomechanical interaction (*44-47*). Theoretical investigations have predicted intriguing quantum revivals (*48*) and quantum persistent tennis racket dynamics (*49*) of nanorotors. In addition, a levitated non-spherical nanoparticle in a vacuum is an ultrasensitive torque detector (*25, 34, 38*) thanks to its high quality factor and small moment of inertia. It is a rare system that can detect the Casimir torque due to quantum vacuum fluctuations (*50, 51*) and the long-sought quantum vacuum friction (*52-*

*56*). However, such applications require a nanoparticle to be driven to rotate near a surface at sub-micrometer separations in a vacuum, which has not been demonstrated yet.

In this article, we optically levitate a nanodumbbell (*36, 38*) at a sub-micrometer separation from a surface for the first time, and drive it to rotate at GHz frequencies near the surface. Despite the small particle-surface separation, we achieve stable GHz rotation of a nanodumbbell near a surface without external feedback cooling. The standing wave formed near the surface due to surface reflection provides stronger spatial confinement than the optical tweezers in free space, which helps to stabilize the trapping. The silica nanodumbbell consists of two nanospheres with diameters of 144 nm. In the first stable trapping well near the surface, which is about 430 nm away from the sapphire surface, we drive a silica nanodumbbell to rotate up to 1.6 GHz in a high vacuum. This mechanical rotation corresponds to a linear speed of 1.4 km $\cdot$ s$^{-1}$ at the tip of the nanodumbbell relative to the surface. Such a record-high relative speed at a sub-micrometer separation will be ideal for detecting the long-sought quantum vacuum friction (*52-56*).

In the experiment, we measure the torque sensitivity and three-dimensional force sensitivity of a nanodumbbell levitated near a sapphire surface. The nanodumbbell near the surface demonstrates a torque sensitivity of $(5.0 \pm 1.1) \times 10^{-26}$ NmHz$^{-1/2}$ at $6.1 \times 10^{-5}$ Torr at room temperature. Compared to the free space case, the nanodumbbell maintains its high torque and force sensitivity when it is levitated near the surface in a vacuum. Thus, a levitated nanoparticle near a surface will be ideal for investigating fundamental particle-surface interactions and other near-field effects. As an example of potential applications, we levitate a nanodumbbell in the first trapping well near a gold nanograting with a stripe width of 300 nm and detect the near-field interference pattern of the laser beyond the optical diffraction limit. In contrast, the effects of the near-field interference is negligible when the nanodumbbell is trapped in the second well that is about 1.2 µm from the nanograting.

In addition, we calculate the Casimir torque on a nanodumbbell levitated near a nanograting, which polarizes nearby quantum vacuum fluctuations (*50, 57, 58*). The calculated Casimir torque can be more than $10^{-24}$ Nm under realistic conditions. Thus, our system will be sensitive enough to measure the Casimir torque. Our work is an important development in levitated optomechanics. It shows a levitated nanodumbbell near a surface in a vacuum will be able to probe surface interactions with ultrahigh torque and force sensitivities that are not achievable with conventional atomic force microscopes.

**GHz rotation near a surface**
In the experiment, we first levitate a silica nanodumbbell with a tightly focused 1550 nm laser in free space (See supplementary Fig. S1 for more details). The laser power at the trapping region is about 200 mW. The nanodummbell is composed of two silica nanospheres with a diameter of 144 nm. With a linearly-polarized trapping laser, we measure the power spectrum density (PSD) of the trapped nanodumbbell at 10 Torr to verify its geometry. The typical PSD of a trapped nanodumbbell shows a large torsional peak. The measured damping rate ratio of the CoM motion perpendicular and parallel to the electric field is close to 1.27 for a nanodumbbell (*36*).

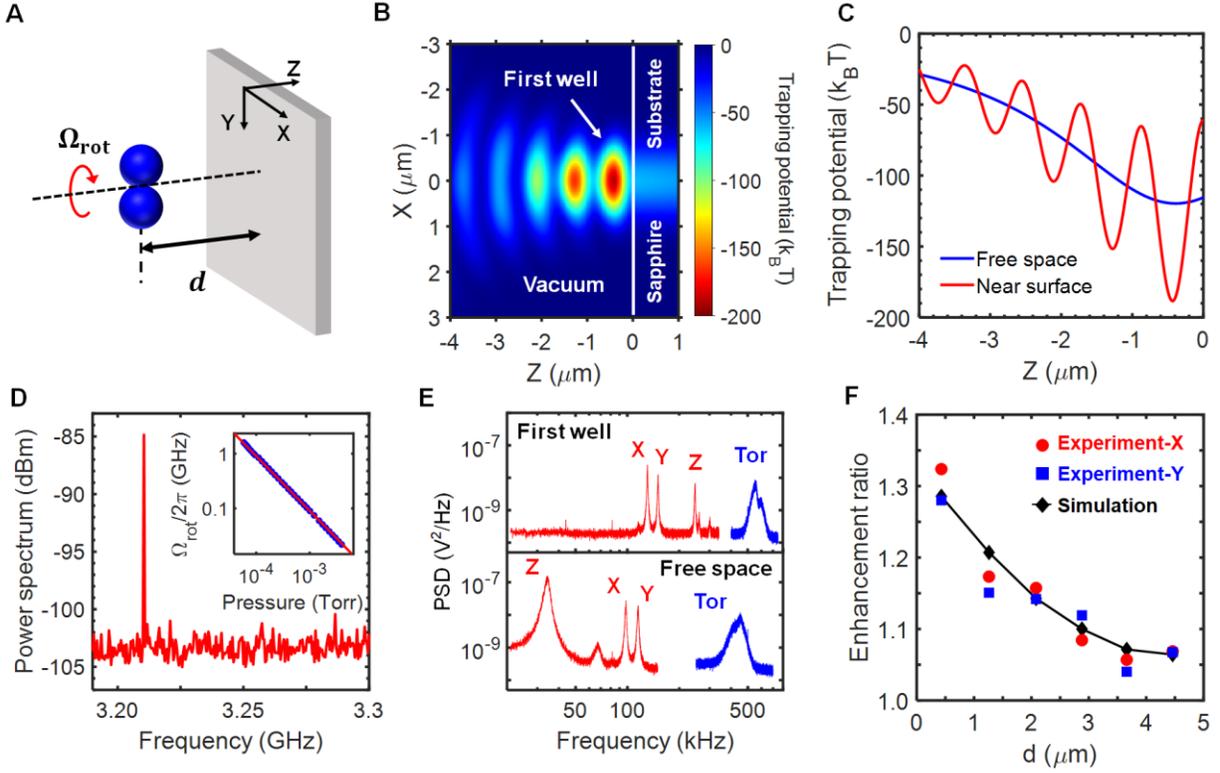

**Fig. 1. Optical levitation and GHz rotation of a nanodumbbell near a sapphire surface. (A)** Schematic of a nanodumbbell rotating at a separation of $d$ from the sapphire surface. **(B)** Simulated trapping potential for a nanodumbbell levitated near the sapphire surface in the unit of $k_BT$ (T = 300 K). Discrete trapping wells are formed due to the standing wave effect near the surface. The trapping position of the first well is about 430 nm from the surface. **(C)** Analytically calculated trapping potential of a nanodumbbell along the propagation direction of the trapping laser. The red and blue curves are trapping potentials near the sapphire surface and in free space, respectively. **(D)** Measured power spectrum of the rotational signal of a nanodumbbell trapped in the first well near the sapphire surface. The peak near 3.2 GHz corresponds to a mechanical rotation frequency of 1.6 GHz at the pressure of $5.9 \times 10^{-5}$ Torr. The inset figure shows the measured rotation frequency as a function of pressure (blue dot) and a fitting curve (solid red line) proportional to $1/P$. **(E)** PSDs of a nanodumbbell trapped in the first well near the sapphire surface (top) and in free space (bottom) at 1.5 Torr. The trapping laser is linearly polarized. The oscillation frequencies of the CoM (Red) and torsional (Blue) motions in the first well are significantly larger than those in free space. **(F)** Enhancement ratios of the trapping frequencies of optical tweezers near a sapphire surface over those in free space as a function of the distance of the trap from the sapphire surface. Red dots and blue squares are measured enhancement ratios for X and Y CoM motions, respectively. The simulation results (black diamonds) are calculated based on the simulated trapping potential shown in (B).

Once the geometry of the nanodumbbell is confirmed in the free space optical trap, we use a high voltage source to create air discharge to neutralize the nanodumbbell (supplementary Fig. S2). After neutralization, a sapphire surface is inserted perpendicularly to the axis of the trapping laser, and the nanodumbbell will be trapped near the sapphire surface (Fig. 1(A)). The partly reflected laser from the surface interferes with the original trapping laser and forms a partial standing wave. This creates discrete trapping wells near the sapphire surface, as shown in Fig. 1(B) and 1(C). The nanodumbbell near the sapphire surface is trapped in one of the trapping wells. By controlling the surface position along the z direction before inserting the sapphire substrate, we can load the nanodumbbell into different trapping wells near the surface. The equilibrium positions of the trapping wells are located at the anti-nodes of the standing waves. The distance between the antinodes and the surface is approximated by $d = (2N − 1)\lambda/4$, where $N = 1,2,3, ...$ is the well number starting from the surface, and $\lambda = 1550$ nm is the wavelength of the trapping laser. From the simulated trapping potential (Fig. 1(B)), the separation between the equilibrium position of the first trapping well and the surface is about 430 nm. It is not exactly $\lambda/4$ because the laser is a focused beam instead of a parallel beam.

After the nanodumbbell is trapped in the first well near the sapphire surface, the polarization of the trapping laser is changed from linear to circular by rotating a quarter waveplate. The angular momentum of the trapping laser applies torque on the nanodumbbell and drives it to rotate in a vacuum. The rotation frequency is determined by the balance of optical torque ($M_o$) and the frictional torque from the residual air given by $M_f = −I\gamma\Omega_{rot}$. Here $I$ is the moment of inertia of the nanodumbbell, $\gamma$ is the rotational damping rate and $\Omega_{rot}$ is the angular velocity of the rotation. While the optical torque is independent of the pressure, the rotational damping rate is proportional to the air pressure $P$. Therefore, the rotation frequency is inversely proportional to the air pressure, as shown in the inset of Fig. 1(D). At $5.9 \times 10^{-5}$ Torr, the mechanical rotation frequency of the nanodumbbell reaches 1.6 GHz (Fig. 1(D)). This corresponds to a linear velocity of $1.4 \text{ km} \cdot \text{s}^{-1}$ between the tip of the nanodumbbell and the sapphire surface, separated by 430nm. While a nanosphere has been levitated near a surface before (*13-15*), to our best knowledge, this is the first report on optical levitation and GHz rotation of a non-spherical nanoparticle near a surface. Such a levitated GHz nanorotor near a surface can be used to explore fundamental physics like measuring quantum vacuum friction (*52-56*).

As a result of the standing wave potential, the CoM motion of a nanodumbbell levitated near the surface is significantly different from free space. Fig. 1(E) shows the PSD of the same nanodumbbell levitated in the first well near the sapphire surface and in free space at 1.5 Torr with the trapping laser linearly polarized. The oscillation frequencies of both CoM and torsional motion are enhanced by the standing wave trapping potential. In particular, the oscillation frequency ($f_z$) of the CoM motion along the *z* direction is enhanced by about 7 times from 35 kHz to 250 kHz. The enhanced trapping frequencies indicate stronger spatial confinement for nanodumbbells levitated near a sapphire surface than in free space. To demonstrate this phenomenon, we record the CoM position of a nanodumbbell levitated in the first well and free space at 1.5 Torr and compare their position distributions. The measured 3D spatial distribution is shown in supplementary Fig. S3. Compared to the free space case, the spatial distribution of the nanodumbbell levitated in the first well is strongly reduced along the *z* direction. The root-mean-square (RMS) value of the *z* position for a nanodumbbell levitated in the first well and free space are 15 nm and 109 nm, respectively. The spatial squeezing along the *z* direction is crucial for stably trapping a nanodumbbell close to a surface, especially at a high vacuum.

Besides the stronger confinement of *z* motion, the oscillation frequencies of *x* and *y* motions are also enhanced by the increased laser intensity gradient due to surface reflection. In the experiment, the *x* and *y* trapping frequencies are recorded when the nanodumbbell is trapped in discrete trapping wells near the sapphire surface and in free space. With that, we calculate the frequency enhancement ratio of *x* and *y* CoM motions due to the surface reflection as a function of the particle-surface separation: $R_{x,y}(d) = f_{x,y}(d)/f_{\text{free space}}$ (shown in Fig. 1(F)). The experimental results agree well with the calculated enhancement ratio using the simulated trapping potential, showing we have indeed trapped the nanodumbbell near the sapphire surface.

**Near-field sensing with a nanodumbbell**

An important motivation for levitating a nanodumbbell near the surface is to study fundamental particle-surface interactions by ultrasensitive torque and force detection. For this purpose, we measure the torque sensitivity and 3D force sensitivity of a levitated nanodumbbell near the sapphire surface. For a nanodumbbell levitated in free space, the torque and force sensitivities are limited by the thermal noise from the residual air in the vacuum chamber and photon shot noise from the trapping laser. When the nanodumbbell is brought close to a surface, its torque and force sensitivities might potentially be affected by the surface.

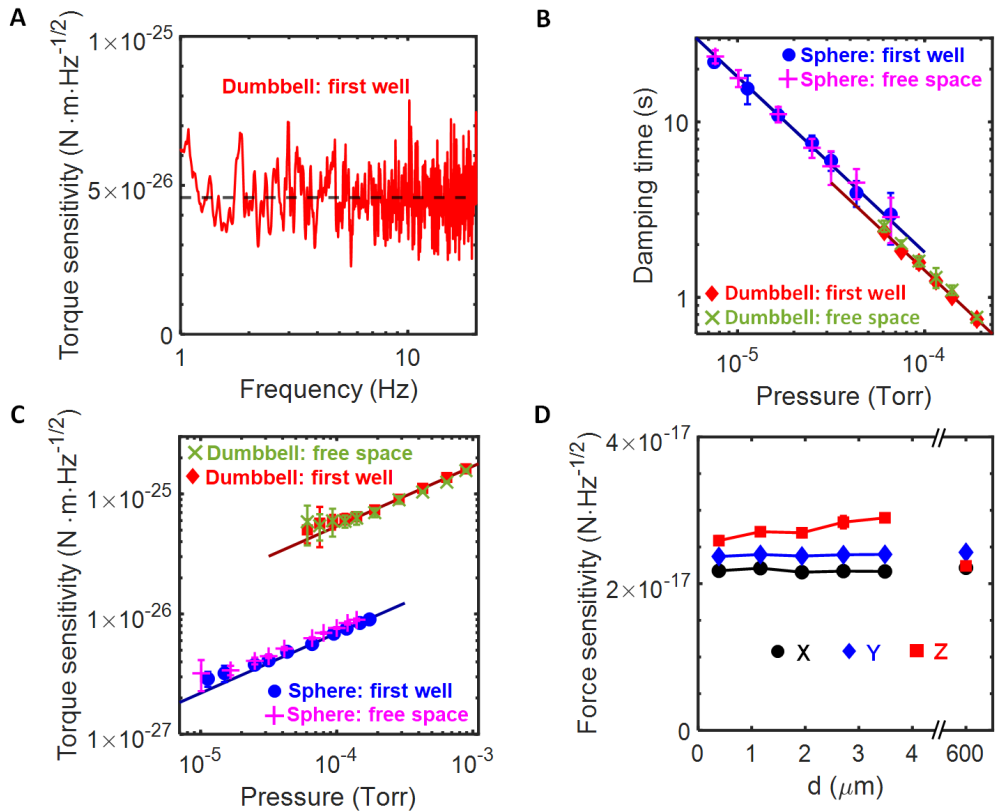

**Fig. 2. Torque and force sensing with a nanoparticle levitated near a sapphire surface.**
(**A**) Torque sensitivity of a nanodumbbell levitated in the first well near a sapphire surface as a function of frequency at $6.1 \times 10^{-5}$ Torr. The dash line shows an average torque sensitivity of $4.6 \times 10^{-26}$ NmHz$^{-1/2}$ for this single measurement. (**B**) Damping time ($\tau$) of a nanodumbbell in the first well (red diamond) and in free space (green cross), a nanosphere levitated in the first well (blue circle), and free space (magenta plus sign). The dark blue and dark red lines are fitting curves based on $\tau \propto 1/P$ for a nanodumbbell and a nanosphere levitated in the first well, respectively. (**C**) The torque sensitivity of a nanodumbbell levitated in the first well (red diamond) and in free space (green cross), a nanosphere levitated in the first well (blue circle), and in free space (magenta plus sign). The dark blue and dark red lines are fitting curves based on $S_T^{1/2} \propto \sqrt{P}$ for a nanodumbbell and a nanosphere levitated in the first well, respectively. (**D**) Three-dimensional force sensitivity of a nanodumbbell as a function of distance to the surface.

Here we measure the torque sensitivity (Fig. 2(A), 2(C)) and rotational damping time (Fig. 2(B)) as a function of pressure for a rotating nanodumbbell and nanosphere levitated in the first well and in free space. The rotational damping time of the nanosphere and nanodumbbell is measured with the ring-up experiment, as shown in supplementary Fig. S4 and Fig. S5. The damping time is inversely proportional to pressure $P$, indicating air damping is the dominant damping source in the measured pressure range. The damping time remains nearly the same when the nanodumbbell is brought from free space to the first well near a surface, indicating no extra damping due to the sapphire surface. If the nanoparticle is trapped in a liquid, the damping rate will increase when it is close to a surface (*59, 60*). In contrast, the damping rate due to residual air molecules at low pressures is insensitive to the separation when the mean free path of molecules is much larger than the size of the nanoparticle and the separation (*61*). The torque sensitivity (*38*) of a nanodumbbell as a function of frequency is $S_T^{1/2}(\Omega_{rot}) = I\sqrt{(\gamma^2 + \Omega_{rot}^2)}\, S_{noise}^{1/2}(\Omega_{rot})$. Here $\gamma = 1/\tau$ is the rotational damping rate. $I$ is the moment of inertia of the nanodumbbell. $S_{noise}^{1/2}(\Omega_{rot})$ is the single-sided PSD of the angular velocity $\Omega_{rot}$ of thermal (Brownian) rotation. The details of determining torque sensitivity are presented in Fig. S4 and Fig. S5 of the supplementary material. For the same nanodumbbell, the difference in the torque sensitivity between the cases near the surface and in free space is within the measurement uncertainty. The torque sensitivities are limited by the air damping and are proportional to $\sqrt{P}$ at high pressure. The torque sensitivity of a nanodumbbell levitated near the surface reaches $(5.0 \pm 1.1) \times 10^{-26}\,\mathrm{NmHz^{-1/2}}$ at $6.1 \times 10^{-5}\,\mathrm{Torr}$. In addition, we perform the same measurement of damping time and torque sensitivity with on a nanosphere, which shows a similar trend as a function of pressure. Compared to a nanodumbbell, a single nanosphere has a smaller momentum of inertia, resulting in a better torque sensitivity than a nanodumbbell at the same pressure. However, an isotropic nanosphere will not experience a Casimir torque that requires asymmetry. Therefore, the nanodumbbell is a better candidate for measuring the Casimir torque than a nanosphere.

The 3D force sensitivity of a nanodumbbell levitated at 1.5 Torr is measured as a function of the separation from the surface (Fig. 2(D)). The force sensitivity (*7*) of the nanodumbbell is limited by thermal noise $S_{F,i}^{1/2} = \sqrt{4k_B T m \Gamma_i}$, where $i = x, y, z$ represents the direction of CoM motion. Here, m is the mass of the nanodumbbell. T = 300 K is the temperature. $\Gamma_i$ is the damping rate of the CoM motion, which is obtained by applying Lorentzian curve fitting to the PSDs of the CoM motion. From the experiment results, the force sensitivity of the nanodumbbell is insensitive to the distance between the nanodumbbell and the surface, in consistent with our results of the torque sensitivity. The average force sensitivity of a nanodumbbell along all three directions is about $2.5 \times 10^{-17}\,\mathrm{N \cdot Hz^{-1/2}}$ at 1.5 Torr. The sensitivity will increase further when the pressure decreases. Thus, a levitated nanodumbbell near a surface can also be used for 3D near-field force microscopy.

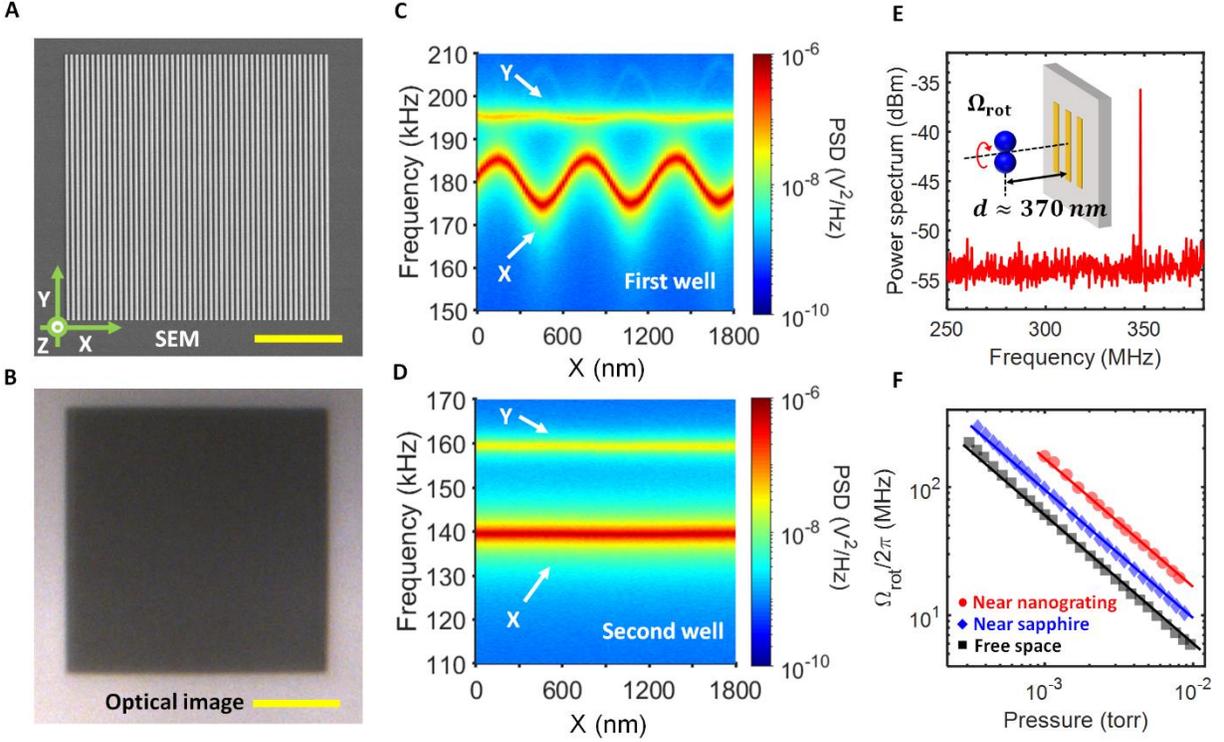

**Fig. 3. Sensing and rotation of a nanodumbbell near a nanograting.** (**A**) SEM and (**B**) optical image of a gold nanograting with a grating period of 600 nm. The width of each gold stripe in the nanograting is 300 nm. The yellow scale bar corresponds to a length of 10 μm. (**C**) and (**D**) are the measured PSDs of a nanodumbbell scanning along the *x* direction when it is trapped in the first and second trapping wells near the nanograting surface, respectively. The frequency change in the first well measures the near-field intensity distribution of the nanograting. (**E**) Power spectrum of the rotation of a nanodumbbell near the nanograting. Inset is the schematic diagram. The peak at 350 MHz corresponds to a rotation frequency of 175 MHz at $1.0 \times 10^{-3}$ Torr. (**F**) Rotation frequency of the nanodumbbell as a function of air pressure when it is levitated in the first well near the nanograting (red dot), in the first well near a flat sapphire surface (blue diamond) and in free space (black square). The solid lines are fitting curves proportional to $1/P$, where *P* is the pressure.

In addition to trapping near a flat sapphire surface, we also levitate a nanodumbbell near a gold nanograting and probe subwavelength light field near the nanograting, as shown in Fig. 3. The nanograting is composed of periodical gold strips on a sapphire substrate (Fig. 3A). The period of the nanograting is $\Lambda = 600$ nm, and the width of each gold stripe is $W_{\text{stripe}} = \Lambda/2 = 300$ nm. The stripes of the nanograting cannot be observed from the optical image shown in Fig. 3(B) as their feature size is smaller than the diffraction limit of our optical imaging system. When the 1550 nm laser is focused on the nanograting (supplementary Fig. S6), the reflected laser from adjacent gold stripes interferes and changes the laser intensity distribution along the *x* direction. As the nanograting period is much smaller than the trapping laser wavelength, such interference only exists in the near-field region and cannot be detected with far-field detectors. The CoM motion of a levitated nanodumbbell is sensitive to the laser distribution near the trapping region. To study the near-field diffraction of the nanograting, we scan the nanograting with a nanodumbbell trapped in the first well near the nanograting.

Fig. 3(C) and Fig. 3(D) are the measured PSDs as a function of scanning position for a nanodumbbell levitated in the first and second wells near the nanograting, respectively. The oscillation frequency of $x$ motion changes periodically when the nanodumbbell scans in the first well, while the frequency remains nearly constant when the nanodumbbell scans in the second well. This shows that the near-field interference disappears at a separation of 1.2 μm from the nanograting surface.

After trapping a nanodumbbell in the first well near the nanograting, we drive it to rotate in a vacuum with the circularly polarized trapping laser. At $1.0 \times 10^{-3}$ Torr, the rotational frequency of the nanodumbbell levitated near nanograting reaches 175 MHz. Fig. 3(F) shows the rotation frequency as a function of pressure for the same nanodumbbell trapped in the first well near a flat sapphire surface, in the first well near nanograting, and in free space. For all three cases, the rotation frequency follows the $1/P$ dependence, as air damping is the dominant damping source. The difference between the rotation frequencies for different configurations at the same pressure is caused by different amounts of reflection from the surface, which changes the optical torque. The nanodumbbell levitated near the nanograting has the highest rotation frequency among these three situations at a given pressure due to the high reflectivity of the gold nanograting. In addition, we have levitated a nanodumbbell near a gold micro-disk (supplementary Fig. S7), which shows an even larger enhancement in trapping frequencies.

**Simulation of the Casimir torque on a nanodumbbell near a nanograting**
According to quantum electrodynamics, a vacuum is not empty but full of virtual photons (quantum vacuum fluctuations). The quantum vacuum fluctuations can lead to an attractive force between neutral plates in vacuum, known as the Casimir force (*62*). If the plates are optically anisotropic, vacuum fluctuations can also induce a torque between them (*63, 64*). While the Casimir force has been measured extensively (*65-70*), the Casimir torque has only been measured with liquid crystals at small separations (*51*). A nanodumbbell levitated near a nanograting will provide an opportunity to study the Casimir torque at different separations systematically (*50, 71*), especially at separations when the retardation effect is important. A nanograting not only causes near-field interference of real photons but also breaks the rotational symmetry of virtual photons near it (*57, 68, 71*). Thus, the Casimir energy between a levitated nanodumbbell and the nanograting is angular dependent, producing a Casimir torque along the rotation axis. Due to the complex geometry, the Casimir torque between a levitated nanodumbbell and a nanostructure has not been calculated before.

Here we numerically calculate the Casimir effect between a nanodumbbell and a gold nanograting (Fig. 4(A)). In particular, we use the SCUFF-CAS3D codes (see Supplementary Materials for more information) (*72, 73*), which are based on the fluctuation-surface current method, to calculate the Casimir effect on a nanodumbbell levitated 370 nm from the nanograting. Compared to a birefringent crystal, the properties of a nanograting can be tuned at will by nanofabrication. The nanograting can affect both spin and spatial distribution of vacuum fluctuations, inducing novel effects that do not exist for natural birefringent crystals (*57, 68*). Our calculated Casimir torque on the nanodumbbell shows a strong dependence on the width of the nanograting, as shown in Fig. 4(B). At the separation of 370 nm, the maximum Casimir torque happens when the grating width is about 300 nm.

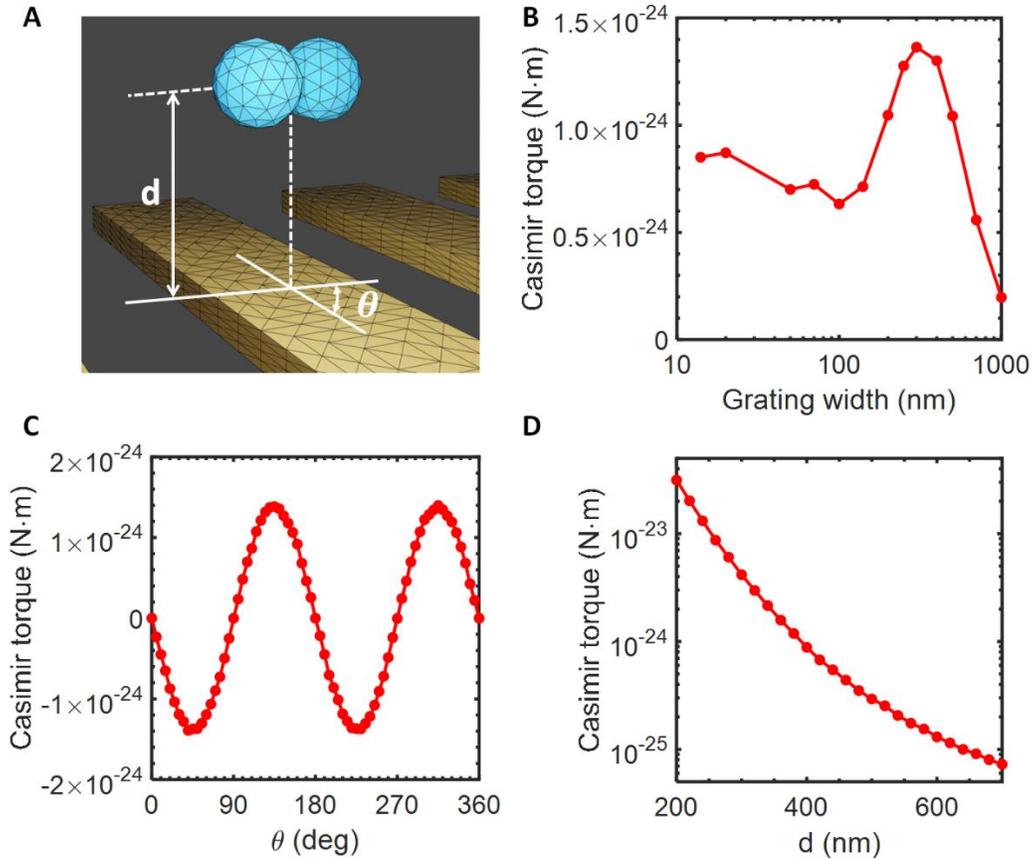

**Fig. 4. Simulation of the Casimir torque on a nanodumbbell levitated near the nanograting.** **(A)** Finite element mesh plot of a nanodumbbell levitated near nanograting used in the simulation. d is the distance between the center of nanodumbbell and the surface of the nanograting. θ is the relative angle between a gold strip in the nanograting and the long axis of the nanodumbbell. **(B)** Casimir torque as a function of grating width in the case of d = 370 nm and θ = 135 deg. Casimir torque on the nanodumbbell near the nanograting (gold stripe width = 300 nm) as a function of **(C)** θ and **(D)** d.

Figure 4(C) shows the calculated angular dependence of the Casimir torque on the nanodumbbell near a nanograting. As a result of the rotational symmetry, the Casimir torque shows a period of 180° over the angle θ between the long axis of the nanodumbbell and the nanograting. The Casimir torque on the nanodumbbell reaches a maximum value of $1.4 \times 10^{-24}$ N · m at $|\theta| = 135°, 315°$. Such a Casimir torque is one order larger than our measured torque sensitivity of a nanodumbbell (Fig. 2(A)). Casimir torque as a function of separation is also calculated in Fig. 4D. The Casimir force on the nanodumbbell at 370 nm separation is about $3.0 \times 10^{-16}$ N (supplementary Fig. S8), which is also well above the force sensitivity of the nanodumbbell. Therefore, this system is sensitive enough to detect the Casimir torque and Casimir force on the nanodumbbell. A free-fall experiment could be used to detect such a static effect (*50, 74*).

**Discussion**

In summary, we have demonstrated near-field GHz mechanical rotation with a nanodumbbell levitated by optical tweezers at about 430 nm from a sapphire surface. The standing wave formed by the reflection of the optical tweezers from the surface provides a stable trapping potential near the surface, where a nanodumbbell is stably trapped and driven to rotate in a high vacuum without feedback cooling. The nanodumbbell maintains its superior torque and force sensitivities at sub-micrometer separation from a surface. The achieved torque sensitivity of a nanodumbbell levitated near the surface is $5.0 \times 10^{-26}$ Nm/Hz$^{1/2}$ at $6.1 \times 10^{-5}$ Torr, and the achieved force sensitivity is about $2.5 \times 10^{-17}$ N·Hz$^{-1/2}$ at 1.5 Torr. We also trap a nanodumbbell near a gold nanograting and use it to detect the near-field laser intensity distribution beyond the diffraction limit. Our simulation shows this system is sensitive enough to detect the Casimir torque on a nanodumbbell near a nanograting.

Currently, the highest rotation frequency of our nanodumbbell near a surface is limited by the increased CoM vibration amplitude at low pressure. In the future, external feedback cooling can be used to stabilize the CoM motion to achieve a higher rotational frequency limited by the ultimate tensile strength of the material (*39*). To detect the Casimir torque with a levitated nanodumbbell, we can use the free-fall method (*50, 74*). The prerequisite of such measurement is to cool the CoM motion as well as the torsional motion of the nanodumbbell simultaneously. This can be done with feedback cooling (*40, 41*) or coherent scattering (*42*). The nanodumbbell can detect both Casimir torque and force in the same free fall experiment near the nanograting. A nanoparticle rotating at high speed near a surface will also be ideal for detecting the quantum vacuum friction (*52-56*) at a high vacuum when the effect of air damping is smaller than the quantum vacuum friction (*38*).

**Acknowledgments**
We thank Alejandro J. Grine, Francis Robicheaux and Jonghoon Ahn for helpful discussions. We acknowledge the support from the Office of Naval Research under Grant No. N00014-18-1-2371 and National Science Foundation under Grant PHY-2110591. This project is also partially supported by the Laboratory Directed Research and Development program at Sandia National Laboratories, a multimission laboratory managed and operated by National Technology and Engineering Solutions of Sandia LLC, a wholly owned subsidiary of Honeywell International Inc., for the U.S. Department of Energy's National Nuclear Security Administration under Contract No. DE-NA0003525. This paper describes objective technical results and analysis. Any subjective views or opinions that might be expressed in the paper do not necessarily represent the views of the U.S. Department of Energy or the United States Government.